\documentclass[pre,superscriptaddress,twocolumn]{revtex4}

\usepackage{graphicx}
\usepackage{dcolumn}
\usepackage{bm}

\usepackage{amssymb}
\usepackage{amsmath}
\usepackage{amsthm}
\usepackage{bbm}

\begin{document}


\def \Z {\mathbb{Z}}
\def \R {\mathbb{R}}
\def \C {\mathbb{C}}
\def \La {\Lambda}
\def \la {\lambda}
\def \ck {l}

\newtheorem{theorem}{Theorem}
\newtheorem{lemma}{Lemma}
\newtheorem{corollary}{Corollary}

\title{Convergence radius of perturbative Lindblad driven non-equilibrium steady states}

\author{Humberto C. F. Lemos}
\email[Corresponding author: ]{humbertolemos@ufsj.edu.br}
\affiliation{Faculty of Mathematics and Physics, University of Ljubljana, Jadranska 19, SI-1000 Ljubljana, Slovenia}
\affiliation{Departamento de F\'{\i}sica e Matem\'atica, CAP -
Universidade Federal de S\~ao Jo\~ao del-Rei, 36.420-000, Ouro Branco, MG, Brazil}
\author{Toma\v{z} Prosen}
\email{tomaz.prosen@fmf.uni-lj.si}
\affiliation{Faculty of Mathematics and Physics, University of Ljubljana, Jadranska 19, SI-1000 Ljubljana, Slovenia}

\begin{abstract}
We address the problem of analyzing the radius of convergence of perturbative expansion of non-equilibrium steady states of Lindblad driven spin chains. A simple formal approach is developed for systematically computing the perturbative expansion of small driven systems. We consider the paradigmatic model of an open $XXZ$ spin 1/2 chain with boundary supported ultralocal Lindblad dissipators and treat two different perturbative cases: (i) expansion in system-bath coupling parameter and (ii) expansion in driving (bias) parameter. In the first case (i) we find that the radius of convergence quickly shrinks with increasing the system size, while in the second case (ii) we find that the convergence radius is always larger than $1$, and in particular it approaches $1$ from above as we change the anisotropy from easy plane ($XY$) to easy axis (Ising) regime.
\end{abstract}

\maketitle

\section{Introduction} \label{sec:intro}

Open quantum systems approach \cite{BreuerPetruccione} has generated a great deal of interest in
recent years, in particular since the theory is not only able to efficiently describe subsystems of
large quantum systems in thermal equilibrium, but also captures non-equilibrium physics of systems
driven out of and even far from equilibrium. For example, one can efficiently model the coherent
quantum transport problem in terms of a Lindblad equation with incoherent quantum jump processes
limited to the ends of the chain \cite{Wichterich}. It has been shown that in the case when the bulk dynamics is
completely integrable, one can often write exact solutions for the many-body density matrix of the
boundary driven steady state \cite{TopicalReview} and these, in turn, generate interesting new non-equilibrium
physics, such as quasilocal conservation laws \cite{JSTATreviewIlievskiEtAl}, persistent currents, etc.
Remarkably, this rather formal approach resolved a long lasting debate (starting with the work of
Zotos and collaborators.~\cite{ZotosPrelovsek95}) on existence of anomalous (ballistic) spin
transport at high-temperatures \cite{TP106}, a problem of not only theoretical but also
experimental interest \cite{HHB,HPSRRBBH}.

However, exact analytic solutions for the non-equilibrium steady state are only possible for a
rather limited set of Lindblad dissipators, such as for example, pure magnetic source on one end,
and pure magnetic sink on the other end \cite{TopicalReview}. In cases of more general and generic
boundaries, one may attempt to consider formal perturbative expansions of the steady state in a
parameter which breaks the integrability (solvability).

In this paper, we address the problem of calculating the radius of convergence of such perturbative
expansions for a general finite open quantum system. We consider a specific and widely studied case
of open $XXZ$ spin $1/2$ chain with magnetic pump boundaries which depend on two parameters, (i) the
system-bath coupling parameter, and (ii) the driving (bias) parameter which generates the magnetic
current in the non-equilibrium steady state (NESS).
We demonstrate that, when expanding in the coupling parameter, the case in which the first two orders
are explicitly and analytically known \cite{TP106}, the convergence radius of the series quickly,
probably exponentially, shrinks with increasing the chain length, so the solution may have little
relevance for the correctly scaled thermodynamic limit. On the other hand, if expanding in the
driving parameter, around the so-called {\em linear response regime}, explicit computations strongly
suggest that the radius of convergence remains finite and bounded by $1$ from below, so this
perturbative expansion should be well relevant even in the thermodynamic limit.

The paper is structured as follows. In section \ref{sec:theory} we develop a systematic approach for
calculating the radius of convergence for perturbative expansion of NESS for finite open quantum
systems. 
In sections \ref{sec:epsilon} and \ref{sec:mu} we present numerical results on the example of boundary
driven $XXZ$ chain, for coupling and driving expansions, respectively, and conjecture the asymptotic
behaviors.

\section{Evaluating the radius of convergence} \label{sec:theory}

We study an open anisotropic Heisenberg $XXZ$ spin 1/2-chain with a constant nearest-neighbor
interaction, given by the Hamiltonian
\begin{equation} \label{eq:H}
H = \sum_{j=1}^{N-1} \sigma_{j}^{x} \sigma_{j+1}^{x} + \sigma_{j}^{y} \sigma_{j+1}^{y}
+ \Delta \sigma_{j}^{z} \sigma_{j+1}^{z}\ ,
\end{equation}
where $\sigma_{j}^{x,y,z}$, for $j=1, \ldots , N$ are Pauli operators acting on a Hilbert tensor
product space $\mathcal{H}=(\C^2)^{\otimes N}$. We also use the standard notation
$\sigma^{0}=\mathbbm{1}_{2}$.
The dynamics of the system is given by a Markovian master equation in the Lindblad form
\begin{equation} \label{eq:dyn}
\frac{d \rho(t)}{dt} = -i \left[ H , \rho (t) \right] +
\sum_{k} 2 L_{k} \rho(t) L^{\dagger}_{k} - 
\left\{ L^{\dagger}_{k} L_{k} , \rho(t) \right\}\ ,
\end{equation}
and in this case we have Lindblad operators mimicking magnetic reservoirs acting only at the
boundaries of the chain, namely $L_{1,2} = \sqrt{ \frac{1}{2} \varepsilon (1 \pm \mu) }\ \sigma_{1}^{\pm}$
and  $L_{3,4} = \sqrt{ \frac{1}{2} \varepsilon (1 \mp \mu) }\ \sigma_{N}^{\mp}$. We stress here the
different roles of the two parameters in the Lindblad operators: $\varepsilon$ is the strength of 
coupling of the spin baths at the boundaries of the chain, while $|\mu|\leq 1$ tells us how strong
is the non-equilibrium driving force acting at the edges of the system. If $\mu = 0$, spin baths
acting on $j=1$ and $j=N$ are symmetric, and the system will reach an equilibrium state after a 
sufficient long time (in our case, this will be an infinite temperature Gibbs state -- completely
mixed state). On the other hand, if $\mu = \pm 1$, we have maximal non-equilibrium driving force, and
the analytical solution for the stationary state is well known in this case \cite{TP107,maximal}.

The model is exactly solvable using a Bethe ansatz \cite{Bethe, Bethe2}, but this closed system
solution no longer applies when it is driven far out of equilibrium with a dynamics given by a
Lindblad equation.
In two papers \cite{TP106,TP107}, one of us made a progress and calculated analytically
some physical quantities for the driven $XXZ$ model. In the present paper, we initially address to
\cite{TP106}, where the author formally expanded the solution as a perturbative series in
$\varepsilon$, related to the strength of the coupling between the baths and the boundaries of the
chain, but as it has been already noted there, the ``convergent properties of perturbation series
are unknown".
For sake of understanding, we recall the main steps of such construction. First, the NESS density operator is
given by
$\rho_{\infty} = \lim_{t \to \infty} \rho (t)$, so it is a fixed point for the dynamics
(\ref{eq:dyn})
\begin{equation} \label{eq:NESS}
-i \left( \mathrm{ad} H \right) \rho_{\infty} + \varepsilon \hat{\mathcal{D}} \rho_{\infty} = 0 \ ,
\end{equation}
where $\left( \mathrm{ad} H \right)\rho := \left[ H , \rho \right]$, and the
dissipator here is defined as
\begin{equation} \label{eq:diss}
\hat{\mathcal{D}} := \frac{1}{2} \left( 1 + \mu \right)  \hat{\mathcal{D}}_{+} +
\frac{1}{2} \left( 1 - \mu \right)  \hat{\mathcal{D}}_{-}\ ,
\end{equation}
with
$$
\hat{\mathcal{D}}_{\pm} \rho :=
2 \sigma_1^{\pm} \rho \sigma_1^{\mp} - \left\{ \sigma_1^{\mp} \sigma_1^{\pm} , \rho \right\} +
2 \sigma_N^{\mp} \rho \sigma_N^{\pm} - \left\{ \sigma_N^{\pm} \sigma_N^{\mp} , \rho \right\}\ .
$$
The NESS solution for this regime of weak coupling is then formally expanded as a series in
$\varepsilon$
\begin{equation} \label{eq:series}
\rho_{\infty} = \sum_{n=0}^{\infty} (i \varepsilon)^n \rho^{(n)} =
\rho^{(0)} + \sum_{n=1}^{\infty} (i \varepsilon)^n \rho^{(n)}\ ,
\end{equation}
and na\"{i}vely replacing the series above into Eq.~(\ref{eq:NESS}) we obtain a recurrence relation
\begin{equation} \label{eq:recurr}
\left( \mathrm{ad} H \right) \rho^{(n)} = - \hat{\mathcal{D}} \rho^{(n-1)}, 
\quad \forall n \geq 1,
\end{equation}
which allows us to evaluate the terms of the sequence $\{ \rho^{(n)} \}$ given the initial condition
$\rho^{(0)} = 2^{-N} \mathbbm{1}_{2^N}$.

In \cite{TP106}, the author evaluated the first and second order terms, but that is not the point
here. He raises three questions about some possible mathematical problems in constructing this
sequence:
(i) it is not clear if recurrence relation (\ref{eq:recurr}) should have unique solutions for each
$n$. 
(ii) It is not also clear if it is always possible, for each $n$, to find out the next term
$\rho^{(n+1)}$.
In order to do so, we must prove that $\hat{\mathcal{D}} \rho^{(n)} \in \mathrm{Im} \
(\mathrm{ad} H)$, for any $n$.
(iii) We do not know the radius of convergence for (\ref{eq:series}).
We stress now that we no longer concern about (i) and (ii), since the uniqueness of NESS is
guaranteed by Evans theorem \cite{Evans} -- the application of Evans theorem to the present model is
well discussed in Ref.~\cite{TopicalReview}.
So the only open problem left is (iii).

We start constructing the sequence term by term using the recurrence relation (\ref{eq:recurr}), up
to $\rho^{(n_0)}$, where $n_0$ is the unknown index such that, for the first time, the vector
$\hat{\mathcal{D}} \rho^{(n_0)}$ can be written as a linear combination of the previous ones, i.e.
\begin{equation} \label{eq:CL}
\hat{\mathcal{D}} \rho^{(n_0)} = \sum_{j=1}^{n_0} c_{j}  \Big( \hat{\mathcal{D}} \rho^{(j-1)}\Big) \ .
\end{equation} 
Just to make it clearer, we emphasize now the crucial definition of $n_0$: starting from the first
term of the series (\ref{eq:series}), namely the vector $\rho^{(0)}$, we use recurrence relation
(\ref{eq:recurr}) to find out the next one, $\rho^{(1)}$. Once it was found, we check if vectors
$\hat{\mathcal{D}} \rho^{(0)}$ and $\hat{\mathcal{D}} \rho^{(1)}$ are linearly dependent: if so, we
have found that $n_0=1$ \cite{Kernel}; otherwise we keep using Eq. (\ref{eq:recurr}) again to find
next terms of the series, one by one, until we find a $\rho^{(n_0)}$ such that, as we did on
(\ref{eq:CL}), the vector $\hat{\mathcal{D}} \rho^{(n_0)}$ may be written as a linear combination of
$\hat{\mathcal{D}} \rho^{(0)}, \ldots , \hat{\mathcal{D}} \rho^{(n_0-1)}$, which are linearly
independent by definition.
Of course, once fixed the system size $N$, the Hilbert space is finite with 
$\dim \mathcal{H}=2^{2N}$, so we always will have such index $n_0 \leq 2^{2N} - 1$ -- this minus 1
refers to the first term of the series $\rho^{(0)}$, which is orthogonal to all other vectors 
$\rho^{(n)}$ \cite{Kernel}. 
But $n_0$ is in practice (much) smaller than the dimension of the Hilbert space, we list some values
in section \ref{sec:epsilon}, see table \ref{tab:n0}.
If we manage to find the coefficients $c_j$ of this linear combination above, then we can obtain any
further term of the sequence. We claim that the next term is given by
\begin{equation} \label{eq:next}
\rho^{(n_0+1)} = \sum_{j=1}^{n_0} c_{j}  \rho^{(j)}\ .
\end{equation}
Indeed, if we replace it into the recurrence relation (\ref{eq:recurr}), we get
\begin{eqnarray} \label{eq:solution}
(\mathrm{ad} H) \rho^{(n_0+1)} &=& 
\sum_{j=1}^{n_0} c_j \Big( (\mathrm{ad} H) \rho^{(j)} \Big) =
\nonumber \\
&=& \sum_{j=1}^{n_0} c_j \left( - \hat{\mathcal{D}} \rho^{(j-1)} \right) =
- \hat{\mathcal{D}} \rho^{(n_0)},
\end{eqnarray}
so it is the solution for the recurrence relation in this step. Moreover, we can easily prove that
$\hat{\mathcal{D}} \rho^{(n_0+1)} \in \mathrm{Im}\ (\mathrm{ad} H)$, since the Range set is a
subspace of $\mathcal{H}$, so we are able to keep constructing the sequence term by term.
The next one is found using the same coefficients, 
\begin{equation} \label{eq:next2}
\rho^{(n_0+2)} = \sum_{j=1}^{n_0} c_j \rho^{(j+1)} = \sum_{j=2}^{n_0+1} c_{j-1} \rho^{(j)} \ ,
\end{equation}
and we can easily show that this is the solution following the same steps that we have just used in
Eq. (\ref{eq:next}). 

We could do this repeatedly from now on and get all the terms of (\ref{eq:series}) up to any desired
order $n>n_0$, but the critical step now is to rewrite them as a linear combination of the elements
of the set $\mathcal{B}_{n_0} =\left\{ \rho^{(1)}, \ldots, \rho^{(n_0)}  \right\}$. In other words,
although $\mathcal{B}_{n_0}$ is not a basis for the whole Hilbert space $\mathcal{H}$, the NESS
solution is essentially  in the subspace $\mathcal{H}_{n_0}$ spanned by this set, except for its
zero-th order term $\rho^{(0)}$. Precisely, we have 
$\big( \rho_{\infty}-\rho^{(0)} \big) \in \mathcal{H}_{n_0}$.
We now aim to rewrite each term as a linear combination of these vectors
\begin{equation} \label{eq:seq}
\rho^{(n_0+k)} = \sum_{j=1}^{n_0} c_j \rho^{(j+k-1)}  =: \sum_{j=1}^{n_0} R_{j}^{(n_0+k)} \rho^{(j)} \ ,
\end{equation}
for any $k \geq 1$,
where in the last equality in Eq. (\ref{eq:seq}) above we have just defined the coefficients
$R_{j}^{(n_0+k)}$ for our new $n_0$-dimensional vector $\mathbf{R}^{(n_0+k)}$. 
For $k=1$ is trivial to see that $R_{j}^{(n_0+1)}=c_j$. For $k=2$, we can match equations
(\ref{eq:next2}) and (\ref{eq:next}) to show that
$$
R_{j}^{(n_0+2)} = R_{j-1}^{(n_0+1)} + c_j R_{n_0}^{(n_0+1)}, \qquad \forall\ 1\leq j\leq n_0,
$$
where by convention we have defined $R_{0}^{(n_0+1)}=0$.
Using an analogous procedure, we can straightforward show that the relation above holds for any
$k\in \mathbb{N}$, i.e.
\begin{equation} \label{eq:Rs}
R_{j}^{(n_0+k+1)} = R_{j-1}^{(n_0+k)} + c_j R_{n_0}^{(n_0+k)}, \quad \forall\ 1\leq j\leq n_0,
\forall k \geq 1,
\end{equation}
where once more we have defined, by convention, $R_{0}^{(n_0+k)}=0$.


We can now compactly rewrite it for $n \geq n_0$ as $\mathbf{R}^{(n+1)} = M \mathbf{R}^{(n)}$, if we
define $M$ as the $n_0\times n_0$ square matrix below
\begin{equation} \label{eq:M}
M = \begin{pmatrix}
0 &         &         &   & c_{1}     \\
1 & 0       &         &   & c_{2}     \\
  & \ddots  & \ddots  &   & \vdots    \\
  &         & 1       & 0 & c_{n_0-1} \\
  &         &         & 1 & c_{n_0}   \\
\end{pmatrix}\ ,
\end{equation}
and by induction we can easily get
$$
\mathbf{R}^{(n_0+k)} = M^{k} \mathbf{R}^{(n_0)}, \quad \forall k \geq 1 .
$$
But we can do even more if we apply it backwards: it is obvious to see that in this basis we have
$\mathbf{R}^{(1)}=(1,0,\ldots,0)^T$, and also that $M\mathbf{R}^{(1)}=(0,1,0,\ldots,0)^T=\mathbf{R}^{(2)}$, and so on.
Thus one can rewrite
\begin{equation} \label{eq:rhon}
\mathbf{R}^{(n)} = M^{n-1} \mathbf{R}^{(1)}, \qquad \forall n \geq 1.
\end{equation}
Now we define $\mathbf{R}^{(\infty)}$ as the operator $\big(\rho_{\infty}-\rho^{(0)}\big)$ spanned in
our basis $\mathcal{B}_{n_0}$, and from this we can formally get
\begin{widetext}
\begin{equation} \label{eq:formal}
\mathbf{R}^{(\infty)} = \sum_{n=1}^{\infty} (i \varepsilon)^n \mathbf{R}^{(n)} =
\left(\sum_{n=1}^{\infty} (i \varepsilon)^n M^{n-1}\right) \mathbf{R}^{(1)} =
 i \varepsilon (\mathbbm{1}_{n_0} - i \varepsilon M)^{-1} \mathbf{R}^{(1)},
\end{equation}
\end{widetext}
where the last equality is true if the series in $\varepsilon$ converges.
We have now a very compact and elegant form to express the NESS, and please note that
$\mathbf{R}^{(\infty)}$ gives the components of NESS in the basis $\mathcal{B}_{n_0}$, i.e.
$$
\rho_{\infty} = \rho{(0)} + \sum_{j=1}^{n_0} \mathbf{R}^{(\infty)}_j \rho^{(j)} .
$$
Although on right hand side of (\ref{eq:formal}) we only have $\mathbf{R}^{(1)}$, which is
essentially the operator $\rho^{(1)}$ written in the basis $\mathcal{B}_{n_0}$, we remind that one needs to
construct the sequence $\{ \rho^{(1)}, \ldots, \rho^{(n_0)} \}$ in order to find the coefficients
$c_1, \ldots, c_{n_0}$ defined in Eq. (\ref{eq:CL}) to construct the matrix $M$, and then evaluate
the resolvent $(\mathbbm{1}_{n_0} - i \varepsilon M)^{-1}$. 
But we recap that we have started with a formal series (\ref{eq:series}) which has been rewritten in
a compact form in the last equality of (\ref{eq:formal}).
Therefore the original series converges if the series for the resolvent converges,
and this will happen if 
$\varepsilon<\lambda $, with 
\begin{equation} \label{eq:lambda}
\lambda = \left(\max_{1\leq j\leq n_0} \{ |\lambda_j| \} \right)^{-1},
\end{equation}
 where $\lambda_j$ are the eigenvalues of the
matrix $M$. Now we have an approach to evaluate the radius of convergence $\lambda$ for the series
(\ref{eq:series}).

\section{Weak system-bath coupling regime} \label{sec:epsilon}

The theoretical procedure to find the exact analytic expression (\ref{eq:formal}), described in the
previous section, is quite simple, or better saying, straightforward for small lengths of the
chain $N$.
We used \textit{Mathematica} to perform explicit computations for $N=2,3,4,5$ which give us information
about the radius of convergence $\lambda$ defined by Eq. (\ref{eq:lambda}). 
We emphasize here that all calculations are done in exact arithmetic after setting some values
 for parameters $\mu$ and $\Delta$, but are becoming practically unfeasible for $N\geq 6$.
 Anyway, from the data obtained we could conjecture the behavior for $\lambda$ dependence
on $N$ and $\Delta$.

First of all, the software must compute the critical index $n_0$ defined in equation (\ref{eq:CL}).
We list in Table \ref{tab:n0} some values for $n_0$ in three different perturbation
regimes, and the case (A) corresponds to our first expansion in $\varepsilon$ parameter with
anisotropy $\Delta>0$. 
\begin{table}[h]
\caption{\label{tab:n0}$N$ dependence of the critical index $n_0$ values for different scenarios:
(A) perturbation in $\varepsilon$ parameter, with anisotropy $\Delta>0$,
(B) perturbation in $\varepsilon$ parameter, with anisotropy $\Delta=0$,
and (C) perturbation in $\mu$ parameter, with anisotropy $\Delta>0$.}
\begin{ruledtabular}
\begin{tabular}{cccccc}
$N$   & 2 & 3 & 4  & 5  & 6   \\
$\dim \mathcal{H}=2^{2N}$ & 16 & 64 & 256 & 1024 & 4096 \\
\hline
(A): $\varepsilon$-perturbation, $\Delta>0$ & 2 & 6 & 26 & 98 & N/A \\
(B): $\varepsilon$-perturbation, $\Delta=0$ & 2 & 4 & 6  & 8  & 10  \\
(C): $\mu$-perturbation, $\Delta>0$ & 2 & 4 & 12 & 36 & N/A \\
\end{tabular}
\end{ruledtabular}
\end{table}

We have also checked that the critical index $n_0$ does not depend on parameter $\Delta$, except for
a sharp change at $\Delta=0$ -- see case (B) on table \ref{tab:n0}.
We remind that $\dim \mathcal{H}=2^{2N}$, so this first step shows us that we only need a small number
of vectors on $\mathcal{H}$ to express the NESS solution $\rho_{\infty}$, and this allowed us to
optimize computational resources.
Even so, as previously said, it was not possible to find $n_0$ for a spin chain with size $N\geq 6$,
but anyway we can see some pattern showing up when $N$ varies from 2 to 5 and conjecture how the
radius of convergence $\lambda$ behaves as $N$ grows.

\begin{figure}[h!]
 \includegraphics[width=85mm]{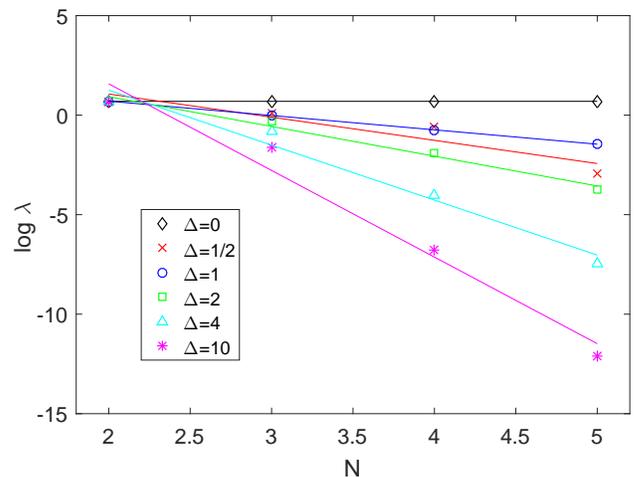}
 \caption{\label{fig:eps_lamb_N} (Color online) Radius of convergence $\lambda$ dependence on system size $N$.
We take $\mu=1/2$ for driving parameter, and different values for anisotropy: 
$\Delta=0$ (black diamonds), $\Delta=1/2$ (red x), $\Delta=1$ (blue circles), $\Delta=2$ (green
squares), $\Delta=4$ (cyan triangles) and $\Delta=10$ (magenta asterisks).
The lines correspond to the best linear fit for $\log \lambda$ vs $N$ in each case.}
 \end{figure}

Once the indices $n_0$ are now known, we can -- by means of \textit{Mathematica} -- find the sequence
$\left\{\rho^{(1)},\ldots,\rho^{(n_0)}\right\}$ using recurrence relation (\ref{eq:recurr}), and
then evaluate the coefficients $c_j$ as in equation (\ref{eq:CL}). 
We fixed $\mu=1/2$ for all the evaluations, and ran the system size $N$ for $2\leq N\leq 5$, and
anisotropy $\Delta>0$ from $10^{-3}$ to $10^{3}$. As described in section \ref{sec:theory}, after
finding the coefficients $c_j$, we construct the matrix $M$ and calculate its spectral radius $\lambda$.
In figure \ref{fig:eps_lamb_N}, we have chosen different fixed values for the anisotropy $\Delta$, and
then we plot a graph of $\log \lambda$ vs $N$, and the results suggest that $\lambda$ decays with
the system size $N$, and apparently faster than exponentially. Moreover, for $\Delta > 1$, the
radius of convergence $\lambda$ decays faster when we increase the anisotropy $\Delta$, as we can
see in Figure \ref{fig:eps_lamb_N}.
When $0<\Delta<1$, $\lambda$ also decays with $N$, as exemplified by data for $\mu=1/2$ plotted
in Figure \ref{fig:eps_lamb_N}. We note that for maximum driving $\mu=1$, the NESS is known
analytically to be a polynomial in $\varepsilon$ of order $2N-2$ \cite{TP107}, so the
$\varepsilon$-expansion has trivially infinite radius of convergence there for any $\Delta$.

To understand what happens to our approach when $\Delta$ gets closer to zero, we used the procedure
described in section \ref{sec:theory} to study the solutions when we have exactly $\Delta=0$ in Eq.
(\ref{eq:H}), namely the $XX$ model.
Repeating all the steps for this simpler model, we are now able to find the $n_0$ indices for system
size $N \leq 6$, and they are listed on case (B) of table \ref{tab:n0}.
Although the solution for $XX$ model is well known \cite{TP107}, we use our approach here to
understand what happens if we keep using smaller positive values for the anisotropy in our
perturbation, trying to get closer to the limit $\Delta \downarrow 0$.
For $\Delta=0$, the values for $n_0$ change drastically, as we can see comparing cases (A) and (B)
in table \ref{tab:n0}.
Except for $N=2$, we can see that $n_0$ is (much) smaller when $\Delta=0$. By the way, these first
values indicate that the critical index increases linearly with the system size, $n_0 = 2(N-1)$.
Anyway, fixing some $N$, our results suggest that the square matrix $M$ is much larger for any
$\Delta>0$, no matter how small the anisotropy parameter is, so $M$ has to dramatically shrink when
we solve the $XX$ model using this approach. We could observe that this affects the behavior of the
inverse of the largest eigenvalue when we increase $N$, this may explain the behavior of $\lambda$
as we approach $\Delta \downarrow 0$ for $N=5$.
Moreover, we were able to find out that for $N$ from 2 up to 6, the matrix $M$ has only two
different eigenvalues, $\pm 1/2$, each one with $(N-1)$ geometric multiplicity. Therefore, when we
increase $N$, the radius of convergence is constant $\lambda_0 = 2$ for $XX$ model.

\section{Linear Response Regime} \label{sec:mu}

In section \ref{sec:theory}, we have described a procedure to study the radius of convergence
$\lambda$ for a perturbation series (\ref{eq:series}) for bath-system coupling parameter
$\varepsilon$. 
We can easily adopt it to study the solutions for the linear response regime, where we use
another parameter as the perturbative one. Here, again we start from the solution in the equilibrium
regime, when we have symmetric magnetic baths ($\mu=0$) coupled to the boundaries. So we would
expect that, at least for small values of the driving force $|\mu|$, we could study the NESS
solution perturbatively. In other words, we are changing the perturbation parameter from
$\varepsilon$ to $\mu$.

We start rewriting the fixed point equation (\ref{eq:NESS}). From (\ref{eq:diss}), after some easy
manipulations, we have
\begin{equation} \label{eq:dissmu}
\varepsilon \hat{\mathcal{D}} = 
\frac{\varepsilon}{2} \left( \hat{\mathcal{D}}_{+} + \hat{\mathcal{D}}_{-} \right) +
\mu \frac{\varepsilon}{2} \left( \hat{\mathcal{D}}_{+} - \hat{\mathcal{D}}_{-} \right) 
=: \hat{\mathcal{D}}_{0} + \mu \hat{\mathcal{D}}_{\mu},
\end{equation}
where in the last equality we defined two new operators, $\hat{\mathcal{D}}_{0}$ and
$\hat{\mathcal{D}}_{\mu}$.
One can easily check that $\hat{\mathcal{D}}_{0}^{\dagger}=\hat{\mathcal{D}}_{0}$.
We get
\begin{equation} \label{eq:NESSmu}
-i \Big[ \left( \mathrm{ad} H \right) + i \hat{\mathcal{D}}_{0} \Big] \rho_{\infty} + 
\mu \hat{\mathcal{D}}_{\mu} \rho_{\infty} = 0 \ ,
\end{equation}
and we define the linear operator $T_{\mu}:=\left( \mathrm{ad} H \right) + i \hat{\mathcal{D}}_{0}$,
just to make expressions simpler. 
In a completely analogous way, we can try a formal series on $\mu$ as solution for NESS 
\begin{equation} \label{eq:seriesmu}
\rho_{\infty} = \sum_{n=0}^{\infty} (i \mu)^n \rho^{(n)}_{\mu} \ ,
\end{equation}
where the terms of the sequence are now labeled with a $\mu$ index just to distinguish that we are
now performing an expansion on this perturbative parameter.
Once again, if we na\"{i}vely substitute it on fixed point equation (\ref{eq:NESSmu}), we find a
similar recurrence relation
\begin{equation} \label{eq:recurrmu}
T_{\mu} \rho^{(n)}_{\mu} = - \hat{\mathcal{D}}_{\mu} \rho^{(n-1)}_{\mu}, 
\quad \forall n \geq 1,
\end{equation}
and again we used the same initial condition $\rho^{(0)}=2^{-N}\mathbbm{1}_{2^N}$, which refers to
the equilibrium solution for $\mu=0$. But at this point we make some comments to show that we have a
more comfortable situation now than in the previous $\varepsilon$-perturbation: one can easily see
that even if $T_{\mu}$ is not a self-adjoint operator, its real part, $\mathrm{ad} H$, as its
imaginary part, $\hat{\mathcal{D}}_{0}$, are both self-adjoint operators. As consequence, we can
prove that $\mathrm{Ker}\ (T_{\mu})$ is a one-dimensional subspace of $\mathcal{H}$ spanned by the
identity vector $\mathbbm{1}_{2^N} = 2^N\rho^{(0)}$. In other words, we no longer have to worry
about degeneracy constructing the elements for the sequence one by one. We also may prove that we
can find any element for the sequence \cite{Evans2}, so we are only concerned about convergence
properties for the series (\ref{eq:seriesmu}). 
In this sense, the perturbation is even simpler for $\mu$ parameter. The rest of the argument follows
exactly the same line as in section \ref{sec:theory}.

From now on, the approach follows as we did on previous section:
First, by means of \textit{Mathematica} code, we use Eq. (\ref{eq:CL}) to obtain the indices $n_0$
for this perturbation, which are listed in the case (C) of table \ref{tab:n0}. 
Again the software could not evaluate it for $N \geq 6$. Then we fixed $\varepsilon=1$ and changed
the system size $N$ from 2 to 5, and anisotropy $\Delta$ from $2^{-14}$ to $2^{14}$. For each case,
we ran the code to evaluate the radius of convergence $\lambda_{\mu}$. Just for the sake of better
understanding, we remind the main steps to the reader: since the index $n_0$  is known for each $N$,
we use recurrence relation (\ref{eq:recurrmu}) to construct the basis $\Big\{ \rho^{(1)}_{\mu} ,
\ldots ,  \rho^{(n_0)}_{\mu}\Big\}$, reminding that the vector $\hat{\mathcal{D}}_{\mu} \rho^{(n_0)}_{\mu}$
can be written as a linear combination of vectors $ \hat{\mathcal{D}}_{\mu} \rho^{(0)}_{\mu} ,
\ldots , \hat{\mathcal{D}}_{\mu} \rho^{(n_0-1)}_{\mu}$, in a completely similar way as we defined it
on Eq. (\ref{eq:CL}). Now we have found coefficients $(c_{\mu})_j$, so we can construct the matrix
$M_{\mu}$, exactly as in (\ref{eq:M}), but with different size $n_0$ and with its last column given
by $(c_{\mu})_1, \ldots ,(c_{\mu})_{n_0}$. Now we can obtain $\mathbf{R}^{(\infty)}_{\mu}$ as in Eq.
(\ref{eq:formal}), and to evaluate the radius of convergence for $\mu$-perturbation as
\begin{equation} \label{eq:lambdamu}
\lambda_{\mu} = \left(\max_{1\leq j\leq n_0} \{ |(\lambda_{\mu})_j| \} \right)^{-1},
\end{equation}
where $(\lambda_{\mu})_j$ are the eigenvalues of the matrix $M_{\mu}$.

\begin{figure}[h!]
 \includegraphics[width=85mm]{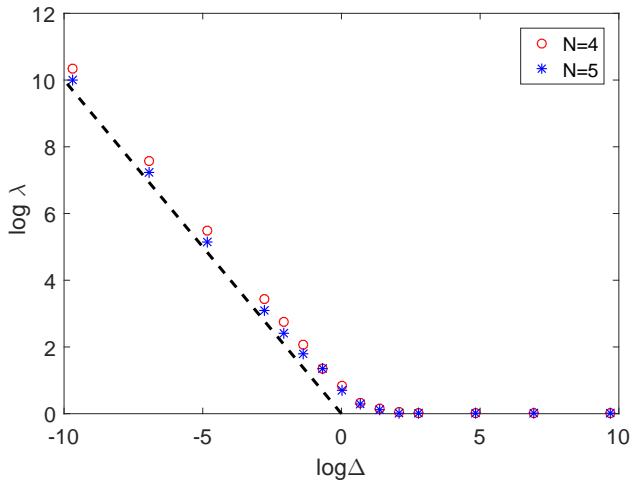}
 \caption{\label{fig:mu_lamb_Delta} (Color online) Radius of convergence $\lambda$ dependence on anisotropy
$\Delta$. We take $\varepsilon=1$ for bath-system coupling strength, and different values for system
size: $N=4$ (red circles) and $N=5$ (blue asterisks). The dashed black line is included as a (-1)
slope guideline.
}
\end{figure}

Here, in the linear response regime, our numerical results allow us to conjecture a very interesting
behavior for $\lambda_{\mu}$. In Figure \ref{fig:mu_lamb_Delta}, for a fixed system size $N$ up to
5, we study how $\lambda_{\mu}$ depends on $\Delta$ in logarithmic scale. We can clearly see a
transition between two behaviors as the black dashed guideline in Figure \ref{fig:mu_lamb_Delta}
indicates: for $0<\Delta<1$, we can see that $\lambda_{\mu}$ decreases as $\Delta$ increases.
However, when we look to $\Delta>1$ we see that the radius of convergence still decreases
as anisotropy increases, but very slightly. Moreover, from our results we can infer a lower bound
$\lambda_{\min}=1$ for the radius of convergence in the linear response regime.
We can see that we always have $\lambda_{\mu}>1$, and it approaches the lower bound when $\Delta$
increases. These results may indicate that the radius of convergence for this perturbation series
does not decay to zero with the system size, in contrast with the $\varepsilon$ perturbation. 
Actually, we conjecture that $\lambda_{\mu} \downarrow 1$ as $\Delta \to \infty$.
If this statement is true, the perturbative solution in this linear response regime is reliable
at least for $\lambda_{\min}=1$, independent of the anisotropy $\Delta$ and the system size $N$.
So, in opposition to the $\varepsilon$-parameter expansion, where the radius of convergence decays
at least exponentially with the system size $N$, here, in linear response regime our results
indicates that in the $\mu$-parameter expansions the radius $\lambda_{\mu}$ remains larger than 1 as
$N$ increases, so the expansion remains relevant, and maybe it is still suitable in thermodynamic
limit.
Moreover, the results indicate that a perturbative solution (\ref{eq:seriesmu}) can always reach the
maximal driving solution $\mu = \pm 1$ regime, for example.

\section{Conclusions}

We have elaborated on a formal and numerical analysis of radius of convergence for the perturbative
solutions of NESS in boundary driven quantum spin chains. We have in particular expanded around the
integrable points, where driven Lindblad equation allows for exact solutions. Even though we could
only do exact numerical computations for relatively short chains (up to 5 sites), our results allow
as to draw some general conclusions. For example, when expanding in the system-bath coupling
strength parameter, the radius of convergence generally shrinks to zero very quickly by increasing
the system size. On the other hand, when expanding in the driving (bias) parameter (the first order
being just the linear response physics), then the radius of convergence appear to be uniformly lower
bounded by $1$.

\begin{acknowledgments}
This work was supported by grant number 249011/2013-1 of CNPq (Brazil), grants P1-0044, N1-0025 of
Slovenian Research Agency, and ERC grant OMNES.
HCFL thanks Alexandre C.L. Almeida for his help and comments.
\end{acknowledgments}



\begin{thebibliography}{15}
\bibitem[{Ker()}]{Kernel}
\bibinfo{note}{It never happens, actually. One may easily note that $\rho^{(0)}
  \in \mathrm{Ker}\, \left( \mathrm{ad} H \right)$.}
\bibitem[{Eva()}]{Evans2}
\bibinfo{note}{Recalling section \ref{sec:theory}, the three points raised
  after recurrence relation (\ref{eq:recurr}): we are claiming here that for
  linear response regime, the points (i) and (ii) are automatically fulfilled.
  Of course, the Evans theorem \cite{Evans} is also valid here.}
\end{thebibliography}

\end{document}